\documentclass[12pt]{JHEP3}
\usepackage{epsfig}
\usepackage{amsmath}
\def\beq{\begin{eqnarray}}
\def\eeq{\end{eqnarray}}
\title{
QCD String as Vortex String in Seiberg-Dual Theory}
\author{
Minoru Eto$^{*,\S,a}$, 
Koji Hashimoto$^{\dagger,b}$ and 
Seiji Terashima$^{\ddagger,c}$\\
${^*}$
{\it INFN, Sezione di Pisa, Largo Pontecorvo, 
3, Ed.~C, 
56127 Pisa,
Italy}\\ 
${^\S}$ {\it Department of Physics,}
\\ 
\hspace{20mm}{\it  University of Pisa, Largo Pontecorvo,
3, Ed.~C, 
56127 Pisa, Italy} \\
${}^\dagger$ {\it Institute of Physics, the University of Tokyo, Komaba,
Tokyo 153-8902, Japan}\\
${}^\ddagger$ {\it Yukawa Institute for Theoretical Physics, \\
\hspace{40mm}Kyoto University, Kyoto 606-8502, Japan}\\
$^a$ E-mail: \email{minoru@df.unipi.it}\\
$^b$ E-mail: \email{koji@hep1.c.u-tokyo.ac.jp}\\
$^c$ E-mail: \email{terasima@yukawa.kyoto-u.ac.jp}\\
}

\abstract{
We construct a classical vortex string solution in a Seiberg-dual theory
of ${\cal N}=1$ supersymmetric $SO(N_c)$ QCD which flows to a confining
phase. We claim that this vortex string is a QCD string, as previouly 
argued by M.~Strassler. In $SO(N_c)$ QCD, it is known that
stable QCD strings exist
even in the presence of dynamical quarks. We show that 
our vortex strings are stable in the Seiberg-dual theory.
}

\preprint{
{\normalsize IFUP-TH/2007-13}\\
{\normalsize UT-Komaba/07-10}\\
{\normalsize YITP-07-34}
}

\begin{document}


\section{Introduction and summary}

Understanding of confinement in QCD is one of the long-standing problems
in particle 
physics. Quarks are confined, and force between them develops a linear
potential described by a confining non-Abelian flux tube, which is
called QCD string. This flux tube 
is expected to be described by dual Meissner effect. In the Meissner
effect, magnetic flux is confined due to condensation of 
electrically charged fields (Higgs fields), which is well described by 
Abelian-Higgs model. The magnetic vortex string solution in that model
is called ANO(Abrikosov-Nielsen-Olesen) vortex \cite{ANO}.  

The duality used in this argument, which mimics the electric-magnetic
duality in Maxwell theory, should be generalized to a non-Abelian
duality. The renowned Seiberg's duality 
\cite{Seiberg:1994pq,Intriligator:1995id} 
provides a proper
basis for addressing this problem, in ${\cal N}=1$
supersymmetric non-Abelian gauge theories. In some situations of the
non-Abelian Seiberg duality, the dual ``magnetic'' theory is weakly
coupled at low energy while the original ``electric'' theory is strongly
coupled, thus classical ANO-like strings may be constructed in the
magnetic theory, as a
concrete realization of the QCD string.  
{\it In this paper, we construct a classical non-BPS 
vortex string solution in a Seiberg-dual of ${\cal N}=1$ 
supersymmetric QCD with non-Abelian gauge groups 
which flows to a supersymmetric confining phase at low energy. }
As for construction of classical non-Abelian vortex strings in the spirit
of studying the QCD strings, see 
\cite{Konishi:2001cj,Auzzi:2003fs,Eto:2006dx,Shifman:2007ce}.

There is an interesting coincidence.
This Seiberg's dual theory which is IR free was used recently \cite{ISS}
to reveal that in fact in ${\cal N}=1$ supersymmetric QCD 
there is a meta-stable supersymmetry-breaking vacua, in addition to the
supersymmetry-preserving vacua.
Here ``QCD'' means 
non-Abelian gauge theories with ``quark'' matter fields in the vector
(fundamental) representation of the gauge group. 
Our previous paper \cite{EHT} studied solitons in these
meta-stable vacua.
In particular, we have shown there that for the dual of the $SU(N_c)$
QCD there is no vortex string, while for the dual of the $SO(N_c)$
gauge theories with $N_f$ flavors (the dual gauge group is $SO(N)$ with
$N=N_f-N_c+4$), there are vortex strings. (For $U(N_c)$ gauge theories,
there are vortex strings associated with the $U(1)$ subgroup.) 
In view of the problem of the QCD strings, is the presence of these
vortex strings in the meta-stable vacua just
a coincidence?  

In fact,
these vortex strings in the
meta-stable vacua are actually directly related to strings in
supersymmetric vacua, in the following way. 
The meta-stable vacua in \cite{ISS}
are obtained when all the quarks obtain masses.
If we tune 
the masses for the $N_f$ quarks in the electric theory as 
$m=(\mu^2, \cdots, \mu^2, 0,\cdots, 0)$ where the first $N$ entries are 
nonzero, the meta-stable vacua restore supersymmetries, as described in
\cite{EHT}. The non-BPS vortex solutions with these tuned mass
parameters in the 
supersymmetric vacua have the same form as those found in \cite{EHT}.
So, vortex strings found in \cite{EHT} are expected to be 
dual of the QCD strings.

In order to identify our classical vortex strings with the
QCD strings, there are two issues. One is the stability, and the other
is the phase. The following are resolution of these issues.
The first issue is that 
in the 
real QCD the QCD strings are unstable. Long strings can break via a pair
creation of a quark and an antiquark, 
so any infinitely long string cannot be
stable, in the presence of dynamical quarks. To evade this difficulty, 
in this paper we consider non-Abelian 
gauge group $SO(N_c)$ instead of usual $SU(3)$
QCD. In $SO(N_c)$ gauge theories, Wilson loops in the spinor
representation can be defined, and associated QCD strings are stable
because they cannot be broken by quarks 
lying in the vector representation.
This is consistent with our findings in \cite{EHT}; only 
for $SO(N_c)$ gauge groups, we found nontrivial topological charges for
the vortex strings. ($U(N_c)$ gauge theories can accommodate strings, but
they are asymptotically non-free because of the crucial $U(1)$ factor
necessary for the vortices to live\footnote{This $U(1)$ is obtained by
gauging the $U(1)_{\rm Baryon}$ global symmetry which is common for
electric and magnetic theories. Consequently, the electric theory has
the same $U(1)$ gauge symmetry and is asymptotically non-free.}.) 
Furthermore, as is well-known, 
Seiberg's duality for $SO(2)$ gauge theory 
with no flavor, $N_f=0$, reduces to the electric-magnetic duality in the
Maxwell theory (the dual group is $SO(2)\sim U(1)$), thus the $SO(N)$ 
series of the duality is not special but naturally shows up.

The other issue is the phase. 
We have to make sure that the electric (original) 
theory is in the confining phase at low energy,
so that the theory actually has the confining QCD strings.
The recipe for this has been studied by M.~Strassler
\cite{Strassler:1997fe} (see also 
\cite{Strassler:1997ny,Strassler:1998fm}) who 
first developed the idea of this identification of the QCD strings
with the vortex strings in Seiberg-dual of $SO(N_c)$ QCD.
Our procedures for a confining phase is as follows. 
First we explicitly construct a
classical vortex solution in the dual $SO(N)$ theory at low
energy, with the quark masses (in the electric theory)
arranged as above. 
The theory is in confining phase at low energy
due to the monopole condensation \cite{Intriligator:1995id}.
Thus our classical vortex string 
can be naturally identified as a QCD string, 
because our solutions are string-like objects which carry
magnetic flux in the theory Seiberg-dual to the confining
gauge theory.
The vortex string solution 
has the tension of the scale $\mu^2$, and 
is reliable for $\mu^2 < \Lambda^2$ where
$\Lambda$ is the scale at which the magnetic $SO(N)$ theory is
strongly-coupled.
The usual QCD string should have the scale of $\Lambda$,
thus accordingly we bring $\mu$
to be large and closer to the scale $\Lambda$.
For large $\mu$, 
$N$ quarks are massive and
decoupled, resulting in the electric theory with
$N_f-N=N_c-4$ flavors whose 
supersymmetric vacuum
is in a confining phase at energy lower than $\Lambda$
\cite{Intriligator:1995id}.
This final procedure, at the same time, brings the tension of our 
vortex string to roughly equal to that of the QCD string.
Note that bringing $\mu$ to the large value requires
a large gauge coupling constant of that energy scale, via
renormalization group. 
There our tree-level analysis of the 
vortex soliton solutions is not valid, but they are 
topologically protected and are expected to remain for large $\mu$.
Our procedure relates the classical vortex strings in the magnetic
theory (which sat at free magnetic phase at $\mu=0$) with the confining
phase (large $\mu$). 

In the following, 
we will present a classical vortex string solution in the Seiberg-dual
of the ${\cal N}=1$ supersymmetric $SO(N_c)$ QCD with $N_f$ quarks in
the vector representation. 
When $N_f = N_c-2$, the dual (magnetic) theory has the gauge group 
$SO(N_f-N_c+4)=SO(2)\sim U(1)$, 
and we will find a vortex string solution as a direct analogue of
the well-known ANO string solution in the Abelian-Higgs model.
This is consistent with the the topological argument of 
\cite{EHT} that the vortex strings have a $Z$ charge.
Our fluctuation analysis will show
that our vortex string is stable classically. 
For generic $N\geq 3$, the topological charge is $Z_2$ as shown in
\cite{EHT}. There we will show that 
a special embedding of the ANO string solution exists. 

\section{$SO(N_c)$ theory and its supersymmetric vacua}

We consider a Seiberg-dual of the $SO(N_c)$ ${\cal N}=1$ supersymmetric
QCD with
$N_f$ quarks in the vector representation of $SO(N_c)$. 
The matter content of the dual magnetic theory for generic
dual gauge group $SO(N)$ with $N=N_f-N_c+4$ is \cite{Seiberg:1994pq}
\begin{center}
\begin{tabular}{c|cccc}
& $SO(N)$ & $SU(N_f)$ & $U(1)'$ & $U(1)_R$ \\
\hline\hline
$\Phi_{[N_f \times N_f]}$ & 1 & $\square\!\square$ & $-2$ & $2$ \\
$\varphi_{[N \times N_f]}$ & $\square$ & $\bar\square$ & $1$ &$0$\\
\end{tabular}
\end{center}
For $\frac32(N_c-2)\geq N_f\geq N_c-2$, the magnetic theory is IR free
and in the so-called free magnetic phase, which we shall make use
of. (For $N_f=N_c-3$ or $N_c-4$, the theory is confining, and so later 
we shall introduce quark mass terms to move from the free magnetic phase
the confining phase.)
The K\"ahler potential, the superpotential and the D-term potential
are
\begin{eqnarray}
K = {\rm Tr} [\varphi^\dagger \varphi] + {\rm Tr} [\Phi^\dagger \Phi],
\;\;
W = h {\rm Tr} \left[\varphi^T \Phi \varphi - M_q \Phi\right],
\;\;
 V_D 
= \frac{g^2}{2}\sum_A \left|
\varphi_i^\dagger T_A \varphi_i
\right|^2.
\label{superpo}
\end{eqnarray}
The symmetric $N_f\times N_f$ matrix $M_q$ is the quark mass matrix in
the electric theory. 
The theory resembles O'Raifeartaigh model, and, in fact, when all the
quarks in the electric theory have the same
non-zero masses, 
\begin{eqnarray}
 M_q = {\rm diag}(\mu^2, \cdots, \mu^2),
\end{eqnarray}
there is a meta-stable vacuum in the magnetic theory. The flavor
symmetry $SU(N_f)\times U(1)'$ is broken down to $O(N_f)$ 
because of the quark mass term.
The vacuum of this theory, meta-stable supersymmetry-breaking one
and the one with supersymmetries dynamically restored, were
studied in detail in \cite{ISS}. 
The meta-stable supersymmetry-breaking vacuum is given by
\begin{eqnarray}
 \Phi = 0, 
\quad
\varphi = \left(
\begin{array}{c}
\varphi_0 \\ 0
\end{array}
\right),
\quad
\mbox{with}
\;\; \varphi_0 = \mu 1_{[N\times N]}.
\end{eqnarray}
The vacuum expectation value $\varphi_0$
gives color-flavor locking.
The vacuum has a cosmological constant, 
$V_{\rm min}= (N_f-N)|h^2 \mu^4|$.

We are interested in supersymmetric vacua which are directly 
accessible from this
meta-stable vacuum, to relate our vortex solutions obtained in
\cite{EHT} with objects in supersymmetric vacua. 
As described in the introduction, and as already
studied in our previous paper \cite{EHT}, if we align the quark masses
in the electric theory as
\begin{eqnarray}
 M_q = {\rm diag}(\mu^2, \cdots, \mu^2, 0,0,0,\cdots,0)
\label{massq}
\end{eqnarray}
in which only the first $N_0$ entries are non-zero with $N_0 \leq N$, 
then the meta-stable vacuum restores
supersymmetries perturbatively, and is identified with the supersymmetric
vacuum of the theory. 
(If $N_0 > N$, perturbative vacua in which our
vortices live are the supersymmetry-breaking meta-stable vacua which are
not of our interest in this paper.)
With this choice of the quark masses, the
``rank condition'' in \cite{ISS} is satisfied, thus the 
cosmological constant of course vanishes.
The supersymmetric vacuum is
\begin{eqnarray}
 \Phi = \left(
\begin{array}{cc}
0 & 0 \\   0 & \Phi_0
\end{array}
\right), \quad
\varphi = \left(
\begin{array}{c}
\varphi_0 \\ 0
\end{array}
\right).
\label{vacuum}
\end{eqnarray}
where $\Phi_0$ is arbitrary constant symmetric matrix with the size 
$(N_f-N_0)\times (N_f-N_0)$, and the diagonal $N \times N$
matrix $\varphi_0$ is 
\begin{eqnarray}
\varphi_0 = {\rm diag}(\mu^2, \cdots, \mu^2, 0, \cdots, 0)
\end{eqnarray}
where the first $N_0$ entries are nonzero.
This gives a color-flavor locking.
Because of the quark mass matrix (\ref{massq}), the flavor symmetry of
the original theory $SU(N_f)\times U(1)'$ is explicitly broken down to 
$O(N_0)\times U(N_f-N_0)$. Therefore the present 
vacuum manifold is quite different from the meta-stable vacuum manifold of
\cite{ISS, EHT}. Our vacuum manifold is just a point\footnote{
Precisely speaking, the vacua consist of two points, 
$Z_2 = O(N_0)/SO(N_0)$.} (times the space spanned by $\Phi_0$), 
and the symmetry
of the vacuum is $SO(N_0)_{\rm C+F}\times G$, where the first 
$SO(N_0)_{\rm C+F}$ is the color-flavor locking symmetry,
and $G \in U(N_f-N_0)$ is the symmetry preserved by $\Phi_0$: for
example if $\Phi_0=0$, $G=U(N_f-N_0)$.
Accordingly, our situation is different from 
\cite{Eto:2007yv} where a ``Seiberg-like'' dual of semilocal 
vortex moduli space
was studied. 
The vacuum has a modulus $\Phi_0$, which survives even in the
limit of large $\mu$ to the confining phase.

In the case of $N=2$ 
in which the magnetic theory has $SO(2)\sim U(1)$ gauge group
and so is in Abelian Coulomb phase, 
the superpotential (\ref{superpo}) is a little modified 
\cite{Intriligator:1995id} as 
\begin{eqnarray}
 W = h \left( a(t)
\sum_{i,j=1}^{N_f}\Phi_{ij} q^+_i q^-_j - \mu^2 \sum_{i=1}^{N_0}
\Phi_{ii}
\right).
\label{so2sup}
\end{eqnarray}
Here $q_i^\pm$ are ``monopoles'' which possess electric $U(1)$ charges
in the dual $SO(2)\sim U(1)$ theory.
In the superpotential, $t \equiv (\det \Phi) /\Lambda^{2(N_c-2)}$ and
$a(0)=1$. 
The mass for the quarks in the electric theory was already chosen as
(\ref{massq}) so that the vacuum is supersymmetric; so we have two
choices, $N_0=1$ or $N_0=2$. 
The superpotential (\ref{so2sup}) looks different from (\ref{superpo}),
but in fact they are very similar to each other. If we redefine the
matter chiral superfields as
\begin{eqnarray}
 q_i^+ = \varphi_i^1 + i \varphi_i^2, 
\quad
 q_i^- = \varphi_i^1 - i \varphi_i^2, 
\end{eqnarray}
where the upper indices are for the $SO(2)$ vector representation, 
then (\ref{so2sup}) reduces to (\ref{superpo}) except for the difference
of the factor $a(t)$. For deriving the vacuum for $N=2$ with the choice
of the quark mass matrix (\ref{massq}) with $N_0=N$,
in fact this factor $a(t)$ is irrelevant, so the supersymmetric 
vacuum configuration is again (\ref{vacuum}).
The dual quarks (which are the ``monopoles'') condense
and the theory is in the Higgs phase with massive photons.

\section{Vortex string solution}

What we have shown in our previous paper \cite{EHT} was that even in
this vacua (\ref{vacuum}) with supersymmetries unbroken, 
there exists a non-BPS vortex string solution, for the case of $U(N_c)$
gauge groups.  
Here we explicitly generalize the study given there to the theory with
$SO(N_c)$ gauge groups, to relate the classical vortex strings with the
QCD strings in confining gauge theories.

The existence of the non-BPS vortex string in the case of $SO(N)$
magnetic theory can be seen in its brane configuration. The $U(N)$
case was studied in our previous paper, and its generalization to the
$SO(N)$ case is straightforward. The brane configuration representing
the vacuum of the $SO(N)$ magnetic theory, derived by using 
the brane realization \cite{Landsteiner:1997vd} of
the Seiberg-duality  in the Hanany-Witten configurations 
\cite{Hanany:1996ie},
was given in
\cite{Franco:2006ht} and shown in the table 
\ref{table:qcd}.\footnote{For
$SU(N_c)$ case and its M-theory lift, see \cite{OO2}.} 
(Note that \cite{Franco:2006ht} 
studied the supersymmetry-breaking meta-stable vacua while
we are interested in the quark mass alignment (\ref{massq}), so all the
D4-branes are parallel to each other in our case, as studied in our
previous paper \cite{EHT} for the $U(N)$ case.) 
As in \cite{EHT}, 
we can add a D2-brane suspended 
between the D4-branes and the
NS5-brane. This D2-brane is oriented along $x^3$ and $x^4$ directions.
This is the vortex string we are interested in.
The orientifold requires that a mirror D2-brane should be added
properly. 
This brane realization of vortices
is along the original idea of \cite{Hanany:2004ea,Lee:1999ze}.

\begin{table}[t]
\begin{center}
\begin{minipage}{130mm}
\begin{center}
\begin{tabular}{c|ccccccccc}
NS 	& 1 & 2 & 3 & -- & -- & -- & -- & 8  & 9 \\
NS'     & 1 & 2 & 3 & 4  & 5  & -- & -- & -- & -- \\
D6 	& 1 & 2 & 3 & -- & -- & -- & 7  & 8  & 9 \\
D4 	& 1 & 2 & 3 & -- & -- & 6  & -- & -- & -- \\
O4 	& 1 & 2 & 3 & -- & -- & 6  & -- & -- & -- \\\hline
D2      & -- & -- & 3 & 4 & -- & --  & -- & -- & -- 
\end{tabular}
\caption{Brane configuration for the magnetic theory with $SO(N)$
gauge group. 
We add a D2-brane (the lowest row) to represent the vortex
 string. } 
\label{table:qcd}
\end{center}
\end{minipage}
\end{center}
\end{table}

So, string theory predicts the existence of a non-BPS vortex string
solution in this magnetic $SO(N)$ theory. 
Being helped by this prediction, we are
able to find an explicit solution of the vortex string.
For the case of $SO(N)$ with $N>2$, the vortex string
solution can be constructed by an embedding of the ANO string into an
$SO(2)$ sub-sector in the $SO(N)$. For the case of $SO(2)$, we will find
that the solution is in fact just a multiple-copied ANO solution. 

For our purpose to show the dual counterpart of the QCD string, 
it is enough to consider one choice of $N$, so let us study the $SO(2)$
case which is the simplest. Furthermore we consider $N_0=N$ for the
quark mass (\ref{massq}). 
Later we study the case of general $N$.
The potential derived from the
superpotential (\ref{so2sup}) and the D-term potential are
\begin{eqnarray}
 V &=& V_F + V_D \nonumber \\
& =& h^2 |a(t)q_1^+q_1^- \!-\! \mu^2|^2
+ h^2 |a(t)q_2^+q_2^- \!-\! \mu^2|^2
+ h^2 \sum_{i=3}^{N_f}|a(t)q_i^+q_i^-|^2
\nonumber \\
&&+ \frac12 h^2 a(t)^2 \sum_{i\neq j}^{N_f}
|q_i^+ q_j^-\! +\! q_j^+ q_i^-|^2 
+ \frac{g^2}{2}\sum_{i=1}^{N_f}
(|q^+_i|^2-|q^-_i|^2)^2 + {\cal O}(\Phi^2).
\label{D-term}
\end{eqnarray}
Here we have omitted writing higher order terms in $\Phi$ because 
this field is kept being the vacuum (\ref{vacuum}) for obtaining the
vortex string solution. So we can take $t=0$ and therefore 
$a(t)=1$.

We work in the convention with the monopoles $q_i^\pm$ rather than
$\varphi_i$, because the former has direct relevance to the ANO solution,
as we will find below. In terms of these monopole fields, the vacuum
(\ref{vacuum}) is
\begin{eqnarray}
 q_1^+ = q_1^-=-i q_2^+ = i q_2^-=\mu, \quad q_{i}^\pm=0 \; (i\geq 3),
\end{eqnarray}
up to the $Z_2 \in O(N_F)$.
It is very natural that the ANO vortex string solution is
embedded in the following manner:
\begin{eqnarray}
&& q_1^+ = (q_1^-)^*=-i q_2^+ = i (q_2^-)^*=f(r)e^{in\theta}, 
\quad q_i^\pm =0 \; (i\geq 3), \nonumber \\
&& A_{\theta} = \frac{-n\alpha(r)}{g}, \quad
A_0=A_3=0.
\label{anosolem}
\end{eqnarray}
Here $r\equiv \sqrt{(x^1)^2 + (x^2)^2}$ and 
$\theta\equiv \arctan (x^2/x^1)$ span the 
cylindrical coordinates with $x^3$. 
One can check that this is in fact a solution of the full system.
The equations for the functions $f(r)$ and $\alpha(r)$ are
\begin{eqnarray}
&& \frac{d^2}{dr^2} f + \frac{1}{r} \frac{d}{dr} f - 
\frac{n^2}{r^2}
 (\alpha-1)^2 f- h^2 (f^2-\mu^2)f=0, 
\\
&& 
\frac{d^2}{dr^2} \alpha - \frac{1}{r} \frac{d}{dr} \alpha 
- 8 g^2 (\alpha-1) f^2= 0.
\end{eqnarray}
\begin{figure}[t]
\begin{center}
 \begin{minipage}{15cm}
\begin{center}
\begin{tabular}{ccc}
\includegraphics[width=16cm]{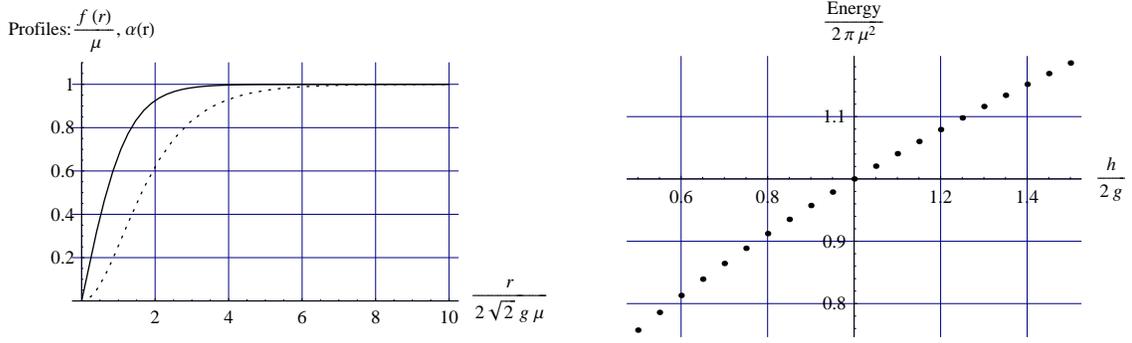}
\end{tabular}
\caption{Left: The functions $f(r)$ (solid line) and
$\alpha(r)$ (dashed line) versus $r$. Right:
 Coupling dependence of the vortex tension. }
\label{evaluation}
\end{center}
 \end{minipage}
\end{center}
\end{figure}
To derive this, we used the K\"ahler potential for the monopole 
fields as $K = (|q^+|^2 + |q^-|^2)$.
The functions $f(r)$ and $\alpha(r)$
interpolate $f(0)=\alpha(0)=0$ and the vacuum with the winding, 
$f(\infty)/\mu=\alpha(\infty)= 1$. 
This is the same as the famous ANO solution \cite{ANO} 
(\cite{Jacobs:1978ch}). See Figure \ref{evaluation}.
The vortex string carries
$n$ unit of the magnetic flux. The width of the vortex measured by the
monopole scalar fields is $\sim h \mu $, while the width of
the concentration of the magnetic flux is $\sim g \mu$. 
When $h=2g$ (which is the BPS limit),
the tension of the vortex string is given by $2 \pi \mu^2$. 
When $h$ differs from the BPS value $2g$ , the tension is 
roughly estimated as ${\cal O}(\mu^2)$.

The embedding ansatz is chosen so that
it does not violate the vanishing of the crossing terms (the fourth and
the fifth terms) in the 
potential (\ref{D-term}).
The winding number is given by $n$, which is the magnetic flux
of the vortex string. This is the dual of the QCD string, for 
the case of $SO(N_c)$ supersymmetric 
QCD with the $N_f = N_c-4$ quarks in the vector
representation of $SO(N_c)$.

In the analysis above, we have chosen $N_0=2$. However, even with
$N_0=1$, the electric theory is confined after the decoupling of this
single massive quark \cite{Intriligator:1995id}, since $N_f=N_c-3$. 
Therefore we expect
that a similar vortex string solution exists also for this $N_0=1$ and
$N=2$. (This case of $N_0=1$ is what M.~Strassler studied in his
original discussion \cite{Strassler:1997fe}.)
Let us present the solution. The potential is 
\begin{eqnarray}
 V & =& h^2 |a(t)q_1^+q_1^- \!-\! \mu^2|^2
+ h^2 \sum_{i=2}^{N_f}|a(t)q_i^+q_i^-|^2
\nonumber \\
&&+ \frac12 h^2 a(t)^2 \sum_{i\neq j}^{N_f}
|q_i^+ q_j^-\! +\! q_j^+ q_i^-|^2 
+ \frac{g^2}{2}\sum_{i=1}^{N_f}
(|q^+_i|^2-|q^-_i|^2)^2 + {\cal O}(\Phi^2).
\label{D-term2}
\end{eqnarray}
The vacuum is unique: $q_1^+=q_1^-=\mu$ and the other fields are zero.
The vortex string solution is
\begin{eqnarray}
q_1^+ = (q_1^-)^* = f(r)e^{in\theta}, \quad A_{\theta}= 
\frac{-n\alpha(r)}{g}, \quad A_0=A_3=0,
\quad q_{i>1}^\pm= 0.
\label{soln01}
\end{eqnarray}
This is the dual of the QCD string, for 
the case of $N_f = N_c-3$.
A similar solution can be easily 
constructed around the massless dyon point 
\cite{Intriligator:1995id} in the moduli space. But this point flows to
a runaway vacuum at low energy, and so the vortex string is irrelevant
to the QCD string.

Next, we study the generic case of $SO(N)$.
The vacuum (\ref{vacuum}) breaks the $SO(N)$ gauge group completely and 
we expect that 
there are non-Abelian strings with the $Z_2(=\pi_1(SO(N)))$ charge.
We choose $N_0=N$ to make sure the confining phase at low energy 
of the electric theory. (One can choose $N_0=N-1$ alternatively,
as in the case of $N=2$ above, but the solution is the same as the one
presented below.)
Without losing generality, we can choose the
embedding of the $SO(2)$ as just the 
first two raws and columns of the $SO(N)$.\footnote{
Similar kinds of
vortex solutions in $SO(N)$ gauge theories have been constructed
and studied in \cite{deVega:1986hm,Konishi:2001cj}.} 
Then, relevant fields have
the potential terms
\begin{eqnarray}
 V_F+V_D & = & h^2 \left|(\varphi_1^a)^2 - \mu^2\right|^2
+ h^2 \left|(\varphi_2^a)^2 - \mu^2\right|^2
+ 4 h^2 \left|\varphi_1^a \varphi_2^a\right|^2 
\nonumber\\
& & + 
\frac{g^2}{8}
\left(
(\varphi_1^a)^* \epsilon_{ab}\varphi_1^b + 
(\varphi_2^a)^* \epsilon_{ab}\varphi_2^b 
\right)^2.
\label{potn3}
\end{eqnarray}
The terms involving the fields $\varphi_i$ with $i\geq 3$ and $\Phi$ are
omitted since they are irrelevant.
For the D-term to be trivially satisfied, we turn on only the real part
of the fields. Then we combine the nontrivial real part of the fields as
\begin{eqnarray}
 \widetilde{\varphi}_1 \equiv 
{\rm Re}\varphi_1^1 + i {\rm Re}\varphi_1^2,
\quad
 \widetilde{\varphi}_2 \equiv 
{\rm Re}\varphi_2^1 + i {\rm Re}\varphi_2^2.
\end{eqnarray}
The $SO(2)$ acts as a $U(1)$ phase gauge rotation on these complex
scalar fields, and therefore the following embedding of the ANO solution
works, 
\begin{eqnarray}
 \widetilde{\varphi}_1 = f(r) e^{in\theta}, \quad
 \widetilde{\varphi}_2 = i f(r) e^{in\theta}.
\label{sonsol}
\end{eqnarray}
The relative phase $i$ in the above embedding is chosen so that
the cross-term $|\varphi^a_1\varphi^a_2|^2$ 
of the F-term potential in (\ref{potn3}) vanishes.
It can be shown straightforwardly that this embedding is a solution
of the whole system, when the other components of the fields are chosen
to be those of the vacuum. 
Note that the solution is the same 
as the solution (\ref{anosolem}) for the $N=2$ case 
though in different notations
and we can write (\ref{sonsol}) as
\begin{eqnarray}
\left( \begin{array}{cc} \varphi_1^1 &  \varphi_1^2 \\ 
\varphi_2^1 &  \varphi_2^2  \end{array} \right)
=f(r) \left( \begin{array}{cc} \cos (n \theta) & \sin (n \theta)  \\ 
-\sin (n \theta) & \cos (n \theta) \end{array} \right).
\end{eqnarray}

We have a choice of how to embed the $SO(2)$ in the whole $SO(N)$. This
should provide an orientational moduli of the vortex string, as in the
famous examples of the $U(N)$ non-Abelian vortex strings 
\cite{Hanany:2004ea,Auzzi:2003fs}. 
The freedom of this choice can be seen in the brane configuration: the 
D2-brane can choose one D4-branes among $N$ of them, to end.\footnote{
The actual
orientational moduli is continuous while the choice of a D4-brane is
discrete: the brane configuration would show only the information of the
Cartan sub-algebra.} In the large $\mu$ limit, this moduli space is
expected to shrink and reduce to a point,
because the confining theory in the electric side 
doesn't know which $N$ one has started with before taking the
limit. 

\section{Stability of the vortex string solution}

The vortex string solution obtained should be stable, 
because at the low energy the electric theory is in the confining phase
and so the flux tube does not decay by broadening itself. 
However, the classical system of the dual theory which admits the vortex
string solutions as above looks similar to the one which admits
so-called semilocal strings \cite{Vachaspati:1991dz}, 
since our dual theory has 
$N_f>N$. It is known that the semilocal strings are unstable and develop
tachyonic instability for a particular parameter region of the
theory.
Actually, the vortex strings in the meta-stable vacua studied in 
\cite{EHT} have such instability.
Here we show that, on the contrary to the expectation from this
similarity, our vortex string solution is stable classically.
Our vortex string is not semilocal\footnote{When $N_0>N$, the 
perturbative vacuum is
supersymmetry-breaking and meta-stable, and the vacuum moduli space is
non-trivial. The vortex strings living there are semilocal, as shown in
\cite{EHT}. The semilocality and its relevance to the confinement 
was discussed in \cite{Shifman:2006kd,Shifman:2007ce}.}, 
and in particular for $N=2$ it has
no moduli space (except for the $\Phi_0$ degree of freedom).

Let us concentrate on the example of $SO(2)$ with $N_0=2$, in which the
solution is given by (\ref{anosolem}). (The system is in fact very
similar to the one derived from ${\cal N}=2$ theory.) 
Fluctuation analysis 
is easier with the following variables
\begin{eqnarray}
 \phi_1 \equiv \frac{1}{\sqrt{2}} \left(
q_1^+ + (q_1^-)^\dagger
\right), \quad
 \widetilde{\phi}_1 \equiv \frac{1}{\sqrt{2}} \left(
q_1^+ - (q_1^-)^\dagger
\right),
\end{eqnarray}
and similar definition for $\phi_2$ and $\widetilde{\phi}_2$ from
$q_2^\pm$. 
The solution lives in the $\phi_1, \phi_2$ 
sector since (\ref{anosolem}) gives
$\widetilde{\phi}_1=\widetilde{\phi}_2=0$.
The solution is stable against fluctuations of $\phi_i$ 
since the analysis is just the same as the Abelian-Higgs model. So,
let us turn on the fluctuation $\widetilde{\phi}_i$.
The potential can be expanded to the second order in
$\widetilde{\phi}_1$ as
\begin{eqnarray}
 \frac{h^2}{4} \left(|\phi_1|^2-2\mu^2\right)^2 + 
\frac{h^2}{4}\left|\phi_1 \widetilde{\phi}_1^\dagger
-\phi_1^\dagger \widetilde{\phi}_1\right|^2 
+ \frac{h^2}{2} \left(2\mu^2 - |\phi_1|^2\right)|\widetilde{\phi}_1|^2
+ \frac{g^2}{2} \left(
\phi_1 \widetilde{\phi}_1^\dagger + \widetilde{\phi}_1\phi_1^\dagger
\right)^2. \nonumber
\end{eqnarray}
Because $f(r)<\mu$ for $r < \infty$,
this is positive semi-definite, and so is the potential for 
fluctuation of $\widetilde{\phi}_2$. The remaining terms relevant 
in the potential (\ref{D-term}) are the third and the 
fourth terms in (\ref{D-term}),
but it is obvious that they are already of the second order in
fluctuations $\widetilde{\phi}_i$ and $q^\pm_{i>2}$, so they are
positive semi-definite. We conclude that our vortex string solution 
(\ref{anosolem}) is stable and has no moduli space
except for massless modes associated with $\Phi_0$. 
The stability of the
solution (\ref{soln01}) can be shown in the same manner.

In our topological argument in \cite{EHT}, there are only $Z_2$ strings
for the case of $N>2$, in contrast to the case of $SO(2)$ where the
winding number $n\in Z$ is the topological charge. We expect that the
$SO(N)$ vortex solutions with higher winding numbers we constructed
are meta-stable in this sense.\footnote{Similar discussions for $SU(N_c)$
QCD and its $Z_{N_c}$strings can be found in 
\cite{Strassler:1997ny,Shifman:2002yi}.}
There may be no topological obstacle to deform the vortex configuration
with a higher winding number to that with a lower one by the 
$Z_2$ grading, but there may be a potential barrier.
The solutions in the case of $SO(2)$ and the 
solutions in $SO(N)$ should be somehow 
related by a mass deformation of the quarks in the original electric
theory. One can change $N$ by changing $N_f$ while $N_c$ being fixed.
In this sense, the ``meta-stable'' solutions (\ref{sonsol}) with higher
winding numbers look rather natural.


\vspace{10mm}
\noindent{\bf Note added:}
While we were writing this paper, we became aware of the paper 
\cite{Shifman:2007kd} which 
discusses relevance of our vortex string found in \cite{EHT} to a QCD
string. 

\acknowledgments 
M.E.~would like to thank M.~Nitta and W.~Vinci for discussions.
K.H.~is grateful to N.~Yokoi for helpful discussions,
and would like to thank Y.~Kikukawa and T.~Yoneya for useful comments.
The work of M.E.~is supported by Japan Society for the Promotion 
of Science under the Post-doctoral Research Program Abroad. 
K.H.~and S.T.~are partly supported by
the Japan Ministry of Education, Culture, Sports, Science and
Technology. 


\end{document}